\documentclass[11pt,twoside]{article}
\usepackage{asp2010}

\resetcounters

\begin{document}

\title{Chromospheric multi-wavelength observations near the solar limb: Techniques and prospects}
\author{C.Beck$^{1,2}$, R.Rezaei$^3$
\affil{$^1$Instituto de Astrof\'{\i}sica de Canarias}
\affil{$^2$Departamento de Astrof{\'i}sica, Universidad de La Laguna}
\affil{$^3$Kiepenheuer-Institut f\"ur Sonnenphysik}}

\begin{abstract}
Observations of chromospheric spectral lines near and beyond the solar limb
provide information on the solar chromosphere without any photospheric
contamination. For ground-based observations near and off the limb with
real-time image correction by adaptive optics (AO), some technical
requirements have to be met, such as an AO lock point that is independent of
the location of the field of view observed by the science instruments, both for 1D and 2D instruments. We show how to obtain simultaneous
AO-corrected spectra in \ion{Ca}{ii} H, H$\alpha$, \ion{Ca}{ii} IR at 854 nm,
and \ion{He}{i} at 1083 nm with the instrumentation at the German Vacuum Tower
Telescope in Iza{\~n}a, Tenerife. We determined the spectral properties of an
active-region macrospicule inside the field of view in the four chromospheric
lines, including its signature in polarization in \ion{He}{i} at 1083 nm. Compared to the line-core intensities, the Doppler shifts of the lines change on a smaller spatial scale in the direction parallel to the limb, suggesting the presence of coherent rotating structures or the passage of upwards propagating helical waves on the surfaces of expanding flux tubes.
\end{abstract}
\section {Introduction}
Spicules are a ubiquitous feature at the solar limb in chromospheric emission lines \citep[see,e.g.,][]{1968SoPh....3..367B,1972ARA&A..10...73B,1998ESASP.421...19S,2000SoPh..196...79S}. Their rapid evolution,
small spatial scales, and low intensity make an observation from the ground
difficult. After initial observations in the 60's and 70's
\citep[e.g.,][]{1962ApJ...136..250Z,1968SoPh....5..131P}, only relatively few
newer observations have been attempted up to today
\citep[][]{1995ApJ...450..411S,2003PNAOJ...7....1M,2005ApJ...619L.195S,2005ApJ...619L.191T,2005A&A...436..325L,2007A&A...472L..51S,2009SoPh..260...59P,2010ApJ...708.1579C,2011A&A...531A.173B}.
Due to the scarcity of the available data, a generic explanation of all
  different types of spicules is missing \citep[but see also][]{2011ApJ...736....9M}. Simultaneous observations in
multiple spectral lines are crucial for determining the physical properties in
spicules such as the temperature or electron density. Different lines provide complementary information through their line widths or emission
  strengths, allowing one to separate for instance between thermal and
non-thermal line width and opacity effects
\citep{1973SoPh...32..345A,2003PNAOJ...7....1M}. Observations of
spicules have been tremendously boosted by the launch of the HINODE satellite,
whose Solar Optical Telescope allows one to observe the solar limb free from
seeing with a broad-band (0.3\,nm width) interference filter centered on the
core of the \ion{Ca}{ii} H line \citep{2007SoPh..243....3K}. These imaging
data revealed the outstanding presence of the spicules near the limb and their
rapid dynamic evolution
\citep[][]{2007PASJ...59S.655D,2008ASPC..397...27S,2010ApJ...714L...1S}, but provide no other information than the total intensity in the filter
  passband for every pixel in a 2D field of view (FOV). This allows one to determine the spatial evolution of spicules and their apparent motions, but makes it nearly impossible to determine the characteristic properties such as densities, mass motions, emissivity, or a possible presence of magnetic fields. 

In this contribution, we present ground-based observations near the solar limb
taken simultaneously in four chromospheric lines (\ion{Ca}{ii} H at
396.85\,nm, H$\alpha$ at 656\,nm, \ion{Ca}{ii} IR at 854\,nm, and \ion{He}{i}
at 1083\,nm) that were obtained in AO-supported observations at the German
Vacuum Tower Telescope \citep[VTT,][]{1985VA.....28..519S} in Iza{\~n}a, Tenerife. Section \ref{setup} explains the requirements for acquiring AO-corrected data near the solar limb and describes the observational setup. Section \ref{general} gives an overview of one of the data sets obtained and investigates the relation between intensity and velocity features. Section \ref{macrospic} discusses the properties of a macrospicule seen inside the FOV. Section \ref{summary} summarizes our findings, whereas the conclusions are given in Sect.~\ref{concl}.
\section{Observational Setup\label{setup}}
\subsection{Requirements for AO observations near the limb}
To  successfully operate an AO system, its wave front sensor (WFS) needs an object image with high contrast. If the images have insufficient
contrast, the wavefront can neither be measured nor corrected accurately. On the solar
disk, the contrast of the granulation pattern reaches about 20\,\% percent
\citep[][]{2010ApJ...723L.154H}, but near the limb the granulation contrast
reduces strongly. One thus needs to find a suitable solar target other than granulation with a high contrast near the limb. Two types of solar features can be used: pores or
sunspots with a local intensity reduction, and faculae with an enhanced intensity. Whereas sunspots or pores are not necessarily found
where needed, suitable faculae can nearly always be encountered. A minor, but
still important issue is that the WFS also should be equipped with a
device for adjusting its light level. The light level near the limb can be
reduced by up to a factor of two relative to the center of the disk. Using a wavelength in the core of a chromospheric line for the WFS might provide a solution to obtaining a high-contrast image beyond the limb, but its feasibility is doubtful: the light level off the limb drops by an order of magnitude to about 1\,--\,3\,\% of the continuum intensity \citep{2011A&A...531A.173B}.

\begin{figure}
\plotone{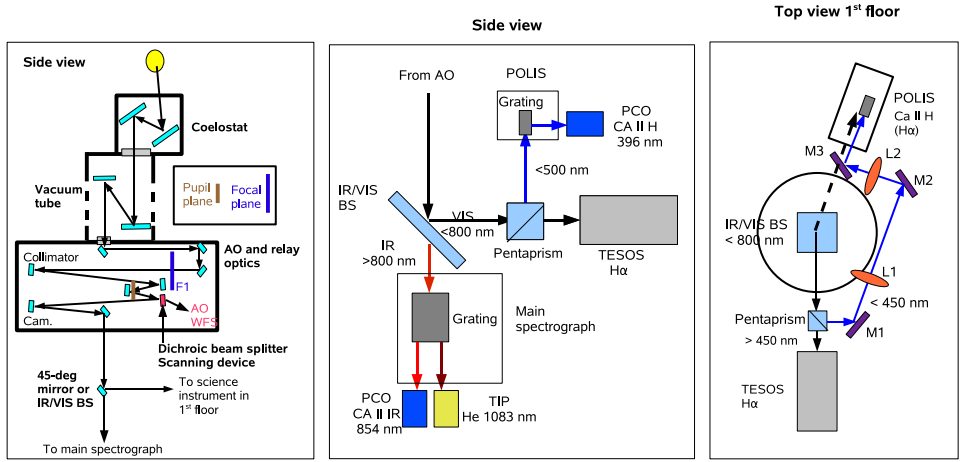} 
\caption{Left: Schematical layout of the VTT. Middle: side view of the instrumentation in first and ground floor. Right: top view of the first floor. The dashed arrow denotes the configuration in 2011 where only POLIS was used.\label{vtt_layout}}
\end{figure}

Now, even if features such as facula provide a high-contrast lock point for a
successful operation of the AO, they also add a requirement: the AO lock point
cannot be changed during the observation. This poses no problem for observations
with 2D spectrometers \citep[e.g.,][]{2010A&A...520A.115B}, but complicates things for observations with slit spectrographs that sequentially scan the solar surface. One possible solution for this problem is achieved in the optical layout of the German VTT sketched in the left panel of Fig.~\ref{vtt_layout}. The AO WFS and the spatial scanning are decoupled here by the fact that the beam splitter (BS), which feeds the AO WFS, is the scanning device at the same time. The AO WFS is fed with the beam that is transmitted through the BS, whereas the post-focus instruments receive the reflected beam. Tilting the BS thus moves the FOV of the science instruments without changing the lock point of the AO WFS. One drawback is that the optimum AO correction is only obtained in a limited radius around the lock point which will then not be centered inside the FOV. 
\subsection{Instrumental setup}
For the simultaneous observations of the four chromospheric spectral lines in
different wavelength regimes, we used two different combinations of the
post-focus instruments at the VTT at that time: the main spectrograph
(\ion{He}{i} 1083\,nm, \ion{Ca}{ii} IR), the Triple Etalon SOlar Spectrometer
\citep[TESOS, H$\alpha$;][]{1998A&A...340..569K,2002SoPh..211...17T}, and the
POlarimetric LIttrow Spectrograph \citep[POLIS, \ion{Ca}{ii} H and
H$\epsilon$, H$\alpha$;][]{2005A&A...437.1159B}. In the following, we will
describe the two different setups used in 2010 and 2011, respectively. POLIS
has been decommissioned by the KIS at the end of 2010, so that only TESOS
remains for observations in the visible range now.

The middle and right panels of Fig.~\ref{vtt_layout} show a sketch of the
light distribution in the two observation runs. Behind the exit of the
adaptive optics' light path, we inserted a dichroic infrared-visible beam
splitter (IR/VIS BS) that splits the incoming light at 800\,nm. For the setup
used in 2010, it reflected the VIS part of the light towards TESOS while
transmitting the IR part towards the main spectrograph. On the main
spectrograph, the Tenerife Infrared Polarimeter
\citep[TIP,][]{2007ASPC..368..611C} was mounted to record spectropolarimetric
data in the wavelength region near 1083\,nm. Next to the cryostat of TIP, we added a PCO 4000 camera to observe the \ion{Ca}{ii} IR line at 854\,nm.  

The VIS part of the light was split once more into a blue ($<$450\,nm) and red
fraction using a dichroic pentaprism close to the entrance of TESOS. The blue
part was folded across the observing room in an one-to-one imaging of the focal plane in front of TESOS onto the focal plane inside POLIS (right panel of Fig.~1). POLIS was not fully operational at that time because of software problems on its camera PCs. We thus mounted a small pick-up mirror behind its grating that reflected the spectrum sidewards out of the instrument towards a second PCO 4000. We used a broad-band interference filter (1\,nm FWHM) centered at 396.85\,nm placed directly in front of the CCD camera to select the order of the spectrum. 

In the setup used in 2011, we reflected the light from the IR/VIS BS directly towards POLIS. Because the POLIS \ion{Ca}{ii} H camera was operational again, we
now used the pick-up mirror inside POLIS to reflect the wavelength range of H$\alpha$ towards a PCO 4000 camera. The slit width corresponded to 0\farcs5 for POLIS and 0\farcs36 for the main spectrograph. 

The data shown here were taken on 12 Apr 2011 between UT 07:49 and 08:14 at the solar position $(x,y)= (-933'',172'')$. The spatial scanning was done in 150 steps of 0\farcs36 with an integration time of 8 seconds per slit position. The spectral coverage for each line, and the spectral and spatial sampling per pixel are listed in Table \ref{tab_spec}. The data were corrected for stray light with an approach similar to that described in \citet{2011A&A...535A.129B}. A measurement of the instrumental PSF was also obtained for an eventual spatial deconvolution to improve the spatial resolution.
\begin{table}
\caption{Wavelength ranges in nm and spectral\,/\,spatial sampling per pixel.\label{tab_spec}}
\centerline{\begin{tabular}{cccc}\noalign{\smallskip}
\tableline
\noalign{\smallskip}
\ion{He}{i} & \ion{Ca}{ii} IR & H$\alpha$ & \ion{Ca}{ii} H \cr
\noalign{\smallskip}
\tableline
\noalign{\smallskip}
1082.34\,--\,1083.45& 853.45\,--\,855.09& 655.12\,--657.04& 396.34\,--\,396.95 \cr
1.0 pm\, /\, 0\farcs18   & 0.8 pm\,/\,0\farcs18 & 2.0 pm\,/\,0\farcs22 & 1.9 pm\,/\,0\farcs30 \cr
\noalign{\smallskip}
\tableline
\end{tabular}}
\end{table}
\begin{figure}
\plotone{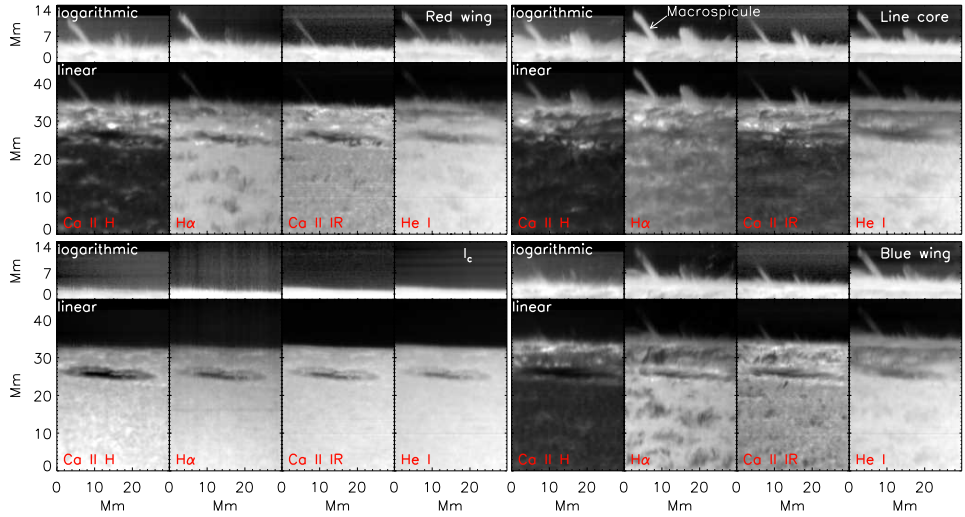}  
\caption{Overview of the observed FOV. Counter-clockwise, starting left bottom: continuum intensity, blue wing, line core, and red wing. The limb region is displayed in a logarithmic intensity scale at the top of each panel. Left to right in each panel: \ion{Ca}{ii} H, H$\alpha$, \ion{Ca}{ii} IR, \ion{He}{i} 1083 nm.\label{fig_overview}} 
\end{figure}
\section{Intensity and velocity structuring \label{general}}
Figure \ref{fig_overview} shows an overview of the FOV as seen in \ion{Ca}{ii} H, H$\alpha$, \ion{Ca}{ii} IR 854.2\,nm, and \ion{He}{i} 1083\,nm. A sunspot located about 5\,Mm from the limb was the AO lock point. The alignment of the FOV in the different wavelengths was only done preliminary without taking into account the temporal variation of the displacements caused by differential refraction \citep[e.g.,][]{2008A&A...479..213B}. The maps of the line-core intensity show various regular spicules near the limb, and a few macrospicules \citep[see, e.g.,][and references therein]{2011A&A...535A..58M}: an isolated macrospicule near the left half of the FOV and a cluster of (macro)spicules near the middle of the FOV. 

The fine-scale structure of the features is more pronounced in the intensity
maps to the blue and red of the line cores taken at wavelengths corresponding to a Doppler shift of about $\pm 15\,$kms$^{-1}$. Especially in H$\alpha$ and \ion{Ca}{ii} IR, individual spatially extended, coherent features in the
  line-core map decompose into multiple strands in the images to the blue and
  red of the line core (e.g., at $x,y = 18$\,Mm, 35\,Mm). This behavior affects both the regular spicules near the limb and the macrospicules, but especially the largest macrospicule marked in Fig.~\ref{fig_overview} shows a pronounced difference. The blue-wing images show mainly the left half of the macrospicule, whereas in the red-wing images the right half is brighter. A similar variation can also be seen still on the disk comparing the three different wavelengths in H$\alpha$. In the line-wing images, bundles of clustered elongated dark streaks appear everywhere in the FOV, whereas the line-core map shows no matching structuring at the same locations. The dark streaks end in locations with enhanced emission in the \ion{Ca}{ii} H line (compare for instance the red line-wing images of H$\alpha$ and \ion{Ca}{ii} H).
\begin{figure}
\centerline{\resizebox{10.cm}{!}{\plotone{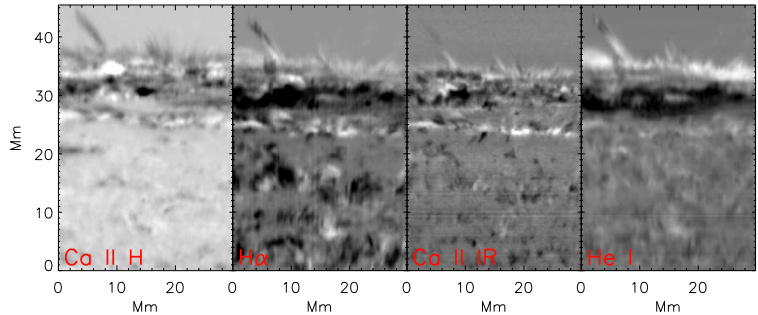}}}
\caption{Intensity difference between red and blue wing.\label{dopplergram}}
\end{figure}

Figure \ref{dopplergram} shows the difference image obtained by subtracting the image to the red of the line core from that to the blue. Especially in \ion{Ca}{ii} IR, the characteristic spatial scale parallel to the limb is smaller in the difference image than in, e.g., the core-intensity map in Fig.~\ref{fig_overview}. The reason why the effects are most pronounced for this line are the line width and the steepness of the profile near the line core. Small Doppler shifts suffice already for a large intensity difference in images taken at a fixed wavelength. The variation of the Doppler shifts along the macrospicule is also very prominent for all lines but \ion{Ca}{ii} H, whose complex line formation makes it less suitable for such an approach to estimate velocities.
\section{Properties of a macrospicule\label{macrospic}}
\begin{figure}
\centerline{\resizebox{12.cm}{!}{\plotone{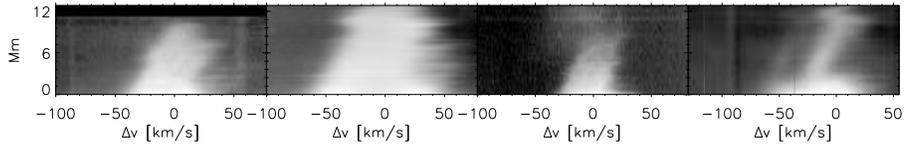}}}
\caption{Intensity profiles along the axis of the macrospicule. Left to right:  \ion{Ca}{ii} H, H$\alpha$,  \ion{Ca}{ii} IR,  \ion{He}{i} 1083 nm. \label{macspic}}
\end{figure}

\begin{figure}
\centerline{\resizebox{9.cm}{!}{\plotone{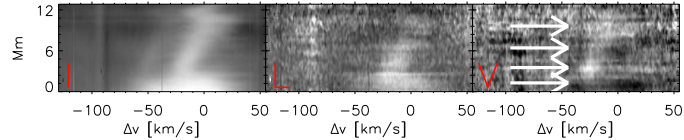}}}
\caption{Polarization profiles in \ion{He}{i} 1083 nm along the axis of the macrospicule. Left to right: Stokes I, total linear polarization $L=\sqrt{Q^2+U^2}$, Stokes $V$.\label{polspic}} 
\end{figure}

The pattern of Doppler shifts in the largest macrospicule is visualized in
more detail in Fig.~\ref{macspic} which shows the intensity spectra
along the central axis of the macrospicule. All four spectral lines clearly
exhibit a smooth transition from blue shifts of about -20 kms$^{-1}$ near the
limb to redshifts at a height of 10 Mm. The location of the absorption
  cores in average profiles was used as zero velocity reference. The polarimetric observations in  \ion{He}{i} 1083 nm with TIP also provided us with the polarization signal of the macrospicule (Fig.~\ref{polspic}). We point out that the interesting fact is already that a polarization signal is seen at a spatial sampling of (0\farcs36)$^2$ with an 8-sec integration time. The macrospicule shows significant linear polarization almost along its full extent. In Stokes $V$, the polarity of the signal seems to flip twice along the length, with an iterative sequence of negative (black) and positive (white) polarization amplitude. This could imply a twisted flux tube, with additional either a constant rotation as required by the Doppler shifts, or a helical wave pattern running along the magnetic field lines. 
\section{Summary and discussion\label{summary}}
From the successful simultaneous observation at the VTT of four of the
strongest chromospheric spectral lines near the solar limb with real-time
correction by adaptive optics, we can derive the following list of requirements for similar observations:
\begin{itemize}
\item Adjustable light level in the AO WFS
\item Independence of AO lock point and scanning procedure
\item Availability of suitable instrumentation
\item Proper light distribution by wavelengths (both setups used had a 100\,\% efficiency for each wavelength range) 
\item Possibility of image rotation for compensation of differential refraction effects.
\end{itemize}
These technical requirements have to be taken into account already in the design phase of a telescope and its instrumentation, because some of them are very difficult (or impossible) to implement afterwards. 

\begin{figure}
\centerline{\resizebox{11.25cm}{!}{\plotone{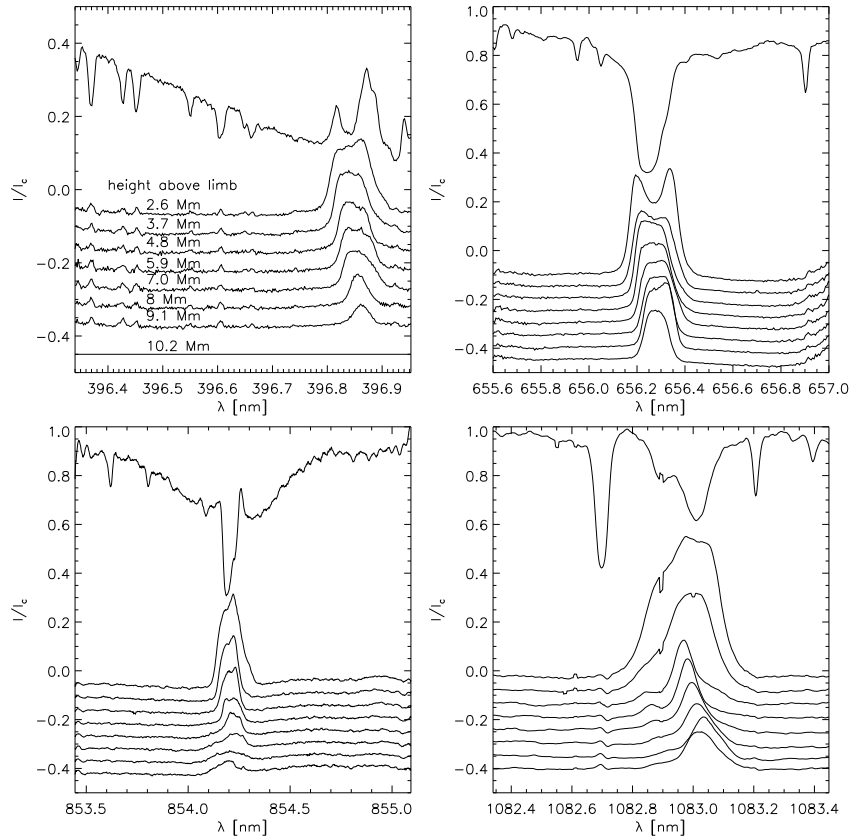}}}
\caption{Individual spectra along the axis of the macrospicule. Clockwise, starting left top:  \ion{Ca}{ii} H, H$\alpha$,  \ion{He}{i} 1083\,nm,  \ion{Ca}{ii} IR 854\,nm. Subsequent spectra were displaced by an arbitrary amount for better visibility. \label{last_fig}}
\end{figure}

For the analysis of the data, we can only outline the prospects of the data at
this stage, using individual co-spatial profiles along the central axis of the
macrospicule as an example (Fig.~\ref{last_fig}). The simultaneous information
of emission amplitude and line width of various lines from the same chemical
element (Ca) gives restrictions on densities and temperature. These
restrictions are even more refined by the other two lines originating from
different chemical elements. The velocity structure can  already be
investigated by any of the lines alone, because they all show similar
patterns. The \ion{He}{i} triplet at 1083\,nm alone provides information on
the thermodynamical structure of the atmosphere by the strength of its two
components, but they also change by coronal EUV radiation
\citep{2008ApJ...677..742C}. Adding the information on the thermodynamics from
the other lines then presents the possibility to determine the coronal
radiation instead. The spectra in Fig.~\ref{last_fig} also provide some information about the optical thickness of the macrospicule: the Ca lines and H$\alpha$ show either a double reversal in the line core (suggestive of self-absorption) or have a flat-topped emission up to a height of about 8 Mm above the limb. This suggests that the macrospicule is optically thick along most of its length.
The macrospicule shows a clear variation of Doppler shifts along its length with a change by about 30 kms$^{-1}$ on about 12\,Mm length. Together with the change of polarity in Stokes $V$ of  \ion{He}{i} at 1083\,nm this suggest a configuration of a twisted flux tube, or a helical wave running along the surface of a flux tube (see \citet{2011A&A...535A..58M} for a different possible scenario). 
\section{Conclusion \label{concl}}
For the future planned 4 m-class telescopes, the current data are very
promising. Polarization signals in the off-limb chromosphere can be detected on
(0\farcs36)$^2$ pixels with 1 m-class telescopes. If one thus does not aim for
diffraction-limited observations, the 4 m-class telescopes should provide a
much better S/N ratio at an improved spatial resolution compared with
the current one.

\acknowledgements The VTT is operated by the Kiepenheuer-Institut f\"ur Sonnenphysik (KIS) at the Spanish Observatorio del Teide of the Instituto de Astrof\'{\i}sica de Canarias (IAC). The POLIS instrument has been a joint development of the High Altitude Observatory (Boulder, USA) and the KIS. C.B~acknowledges partial support from the Spanish Ministerio de Ciencia e Innovaci\'on through project AYA 2010-18029. 
\bibliography{beck}

\end{document}